\newcounter{myctr}
\def\myitem{\refstepcounter{myctr}\bibfont\noindent\ifnum\themyctr>9\else\phantom{0}\fi\hangindent17pt\themyctr.\enskip}

\documentclass{ws-ijqi}
\usepackage{subfigure}
\usepackage{graphicx,color}
\usepackage{bm}

\definecolor{dgreen}{rgb}{0,0.5,0}
\definecolor{delete}{cmyk}{0.5,0,0,0}

\newcommand{\DEL}[1]{{\color{delete}#1}}

\def\bra#1{\langle #1 |}
\def\ket#1{| #1 \rangle}

\newcommand{\tr}{\mathop{\text{Tr}}\nolimits}

\begin{document}
\markboth{P. Facchi \textit{et~al.}} {Phase randomization and typicality in the interference of two condensates}

\catchline{}{}{}{}{}

\title{PHASE RANDOMIZATION AND TYPICALITY IN THE  INTERFERENCE OF TWO CONDENSATES}

\author{PAOLO FACCHI}
\address{Dipartimento di Fisica and MECENAS, Universit\`a di Bari, I-70126 Bari, Italy\\
INFN, Sezione di Bari, I-70126 Bari, Italy}

\author{HIROMICHI NAKAZATO}
\address{Department of Physics, Waseda University, Tokyo 169-8555, Japan}

\author{SAVERIO PASCAZIO}
\address{Dipartimento di Fisica and MECENAS, Universit\`a di Bari, I-70126 Bari, Italy\\
INFN, Sezione di Bari, I-70126 Bari, Italy}

\author{FRANCESCO V. PEPE}
\address{Dipartimento di Fisica and MECENAS, Universit\`a di Bari, I-70126 Bari, Italy\\
INFN, Sezione di Bari, I-70126 Bari, Italy}

\author{GOLAM ALI SEKH}
\address{INFN, Sezione di Bari, I-70126 Bari, Italy}

\author{KAZUYA YUASA}
\address{Department of Physics, Waseda University, Tokyo 169-8555, Japan}

\maketitle

\begin{history}
\received{Day Month Year} \revised{Day Month Year}
\end{history}

\begin{abstract}
Interference is observed when two independent Bose-Einstein
condensates expand and overlap. This phenomenon is typical, in the
sense that the overwhelming majority of wave functions of the
condensates, uniformly sampled out of a suitable portion of the
total Hilbert space, display interference with maximal visibility.
We focus here on the phases of the condensates and their
(pseudo) randomization, which naturally emerges when two-body
scattering processes are considered. Relationship to typicality is discussed and analyzed.
\end{abstract}

\keywords{Interference, Bose-Einstein condensates, typicality, randomness}

\section{Introduction}
\label{sec-intro}

The physical significance of the phase of a Bose-Einstein
condensate (BEC) has raised a number of interesting foundamental
quantum-mechanical
questions.\cite{PW,SB,Leggett,BDZ,PS,PeSm,Leggettbook}
Interference is observed even when two condensates are prepared
\textit{independently}.\cite{exptBEC} This phenomenon,
characteristic of condensates, contrasts with common wisdom on
single-particle double-slit interference
experiments,\cite{Dirac,feynman} where no interference can be
observed unless the relative phase between the two branch waves is
kept constant.\cite{NP93,NPN} In this sense, independent sources
do not interfere (at first order; the second-order Hanbury Brown
and Twiss interference\cite{HBT1,HBT,ref:MandelWolf} is a
different story).

The most credited explanation of the observation of interference
in two-mode Bose systems relies on the beautiful idea that the
relative phase of the condensates is established by measurement.
The phase offset of each single interference pattern changes from
run to run, so that no interference persists if many interference
patterns are superimposed: there is thus no contradiction with the
standard quantum-mechanical interpretation of first-order
interference. This ``measurement-induced interference" was first
proposed in Ref.\ \refcite{JY} and then corroborated by a number
of studies.\cite{CGNZ,WCW,Molmer,CD,Raz} The interpretation
of the experimental results has also been formalized in terms of positive operator valued
measures.\cite{Benatti1,Benatti2}\DEL{.} These ideas bear consequences on our understanding of symmetry breaking
phenomena.\cite{SB,PS,Leggettbook,ref:LeggettSols}

We showed in Ref.\ \refcite{typbec} that interference is robust
with respect to the state preparation. Each time two condensates
are experimentally prepared (e.g.\ out of a single condensate, by
inserting a ``wall'' between
them\cite{exptBEC,Schmied,Schmied2,Schmied3,Schmied4,Schmied5}),
their wave function is sampled out of a portion of the total Hilbert space that depends on experimental
procedures and details (state preparation). Since we have no
access to this information, we have to look at the \emph{typical} features
of such a wave function, namely those features that characterize
its behavior and properties in the overwhelming majority of cases.
We find that the very presence of an interference pattern emerges
as a typical feature of the wave function.\cite{typbec}

In the present article we will build on this observation and focus
on phase randomization effects due to self-interaction within the
condensate, to clarify how the phase randomization process can take place.
This article is organized as follows. In Section
\ref{sec-typicality} we introduce the ensemble of initial states
and review its properties. In particular, we define and analyze
the averages and variances of the physical observables. Section
\ref{sec-observables} is devoted to the study of the general
properties of the observable which are directly related to
interference, in a second-quantization framework. In Section~\ref{sec-gauss} we review the main results on the typicality of
interference between two expanding Gaussian modes, which constitute a realistic model of a BEC interferometry experiment.
In Section~\ref{sec-self} we consider the role of the interaction among particles and write down a simple Hamiltonian model, and
in Section~\ref{sec-typs} we show how an initial coherent state evolves under this Hamiltonian.
Finally, in Section~\ref{sec-pseudo}, we show how this simple physical mechanism
yields a dynamical randomization of the phases in the two-mode Fock
basis.

\section{Distribution of Initial States}
\label{sec-typicality}

We consider a typical experimental setup of BEC interferometry: a
condensate is distributed among two orthogonal modes,
$\psi_a(\bm{r})$ and $\psi_b(\bm{r})$, which are usually spatially
separated at some initial time. Then the atomic clouds are let to
expand, overlap, and (possibly) interfere.

Let us assume that the total number of bosons $N$ is fixed. A
useful basis for such a system is given by two-mode Fock states
\begin{equation}
\label{fock} \ket{ \ell } := \left| \left( \frac{N}{2}+\ell
\right)_a, \left( \frac{N}{2}-\ell \right)_b \right\rangle
= \frac{1}{\sqrt{\left(\frac{N}{2}+\ell\right)! \left(\frac{N}{2}-\ell\right)!}}
({\hat{a}}^{\dagger})^{\frac{N}{2}+\ell} ({\hat{b}}^{\dagger})^{\frac{N}{2}-\ell}
\ket{\Omega},
\end{equation}
with $-N/2 \leq \ell \leq N/2$,
in which the two modes $\psi_a(\bm{r})$ and $\psi_b(\bm{r})$ are orthonormal, and have well-defined occupation numbers. We assume that $N$ is even for simplicity.
The mode operators,
\begin{equation}
\hat{a} = \int d\bm{r}\, \psi_a^*(\bm{r}) \hat{\Psi}(\bm{r}),
\qquad \hat{b} = \int d\bm{r}\, \psi_b^*(\bm{r}) \hat{\Psi}(\bm{r}),
\end{equation}
annihilate the vacuum state $\ket{\Omega}$ and satisfy the canonical commutation relations
$[ \hat{a}, \hat{a}^{\dagger} ] =  [ \hat{b}, \hat{b}^{\dagger} ] = 1$, all the operators of mode $a$ commuting with those of mode $b$.
Here $\hat{\Psi}(\bm{r})$ is the bosonic field operator,
satisfying the canonical commutation relations
$[\hat{\Psi}(\bm{r}),
\hat{\Psi}^\dagger(\bm{r'})]=\delta(\bm{r}-\bm{r}')$, etc.
The number operators $\hat{N}_a={\hat{a}}^{\dagger} \hat{a}$ and
$\hat{N}_b={\hat{b}}^{\dagger} \hat{b}$ count the numbers of
particles in the two modes.

The crucial assumption\cite{typbec}
is that the initial state of the two-mode system is randomly
picked from the subspace spanned by the Fock states with
$|\ell|<n/2$,
\begin{equation}
\mathcal{H}_n=\mathop{\mathrm{span}} \{ \ket{\ell} \; |\;  {-n/2} < \ell < n/2\},
\label{6}
\end{equation}
with $0<n\leq N+1$, where $n$ is odd for simplicity, and the microcanonical density matrix reads
\begin{equation}
\hat{\varrho}_n = \frac{1}{n} \sum_{|\ell| < n/2 }
 \ket{\ell} \bra{\ell}.
 \label{eq:microcanonical}
\end{equation}
The case
$n=1$ was studied by a number of
authors,\cite{JY,CGNZ,WCW,CD,ref:PolkovnikovEPL,ref:Paraoanu-JLTP,ref:Paraoanu-PRA,YI} while
we are more interested in the large-$n$ case. It is not harmful to
think of the ``natural" situation $n = O(\sqrt{N})$, but
we shall work in full generality, with an arbitrary $n=o(N)$.
Surprisingly, interference turns out to be robust\cite{typbec}
against the stronger scaling  $n = o(N)$, which includes, for example, $n \sim N/\log N$.

A general pure state $\ket{\Phi_N}$ of the system drawn from the (unit sphere) of the subspace $\mathcal{H}_n$ can be expanded
in the Fock basis (\ref{fock}) as
\begin{equation}
\ket{\Phi_N} = \sum_{|\ell|<n/2} z_{\ell} \ket{\ell}, \qquad \sum_{|\ell|<n/2} |z_{\ell}|^2=1,
\label{eq:ensemble}
\end{equation}
since $z_{\ell} = 0$ for  $|\ell|\geq n/2$.
The coefficients $z_\ell$ for $|\ell| < n/2$ are randomly
sampled from the  surface of the
$2n$-dimensional unit sphere $\sum_\ell |z_{\ell}|^2=1$.

The assumption of \emph{uniform} sampling is a simplifying
one: the number of states that are actually involved in the
description and their amplitude will depend on the experimental
procedure and the way the two BEC clouds are
created.\cite{esteve} Due to this assumption,
the quadratic \emph{statistical} average over all experimental runs~(\ref{eq:ensemble}) reads
\begin{equation}
\label{psin}
\overline{z_{\ell_1}^* z_{\ell_2}} = \frac{1}{n}
\delta_{\ell_1,\ell_2},
\end{equation}
while the average
of the coefficients themselves, as well as all the quantities that
depend on the phases of the coefficients, will vanish.

In a given run, with state  $\ket{\Phi_N}$,  a quantum observable $\hat{A}$  has expectation value
\begin{equation}
\label{Adef}
A_{\Phi_N}= \bra{\Phi_N} \hat{A} \ket{\Phi_N}.
\end{equation}
Its statistical average over all experimental runs, described by
the uniform ensemble~(\ref{eq:ensemble}), is
\begin{equation}\label{expGEN}
\overline{A} := \overline{ \bra{\Phi_N} \hat{A} \ket{\Phi_N}}
= \tr(\hat{\varrho}_n \hat{A} ) = A_{\varrho_n}.
\end{equation}
There are two distinct fluctuations that characterize a given observable:
i) the \emph{statistical fluctuations} of $A_{\Phi_N}$ over the experimental runs,
quantified by the statistical variance
\begin{equation}\label{1}
(\delta A)^2 := \overline{ A^2 } - \overline{A}^2 = \overline{
\bra{\Phi_N} \hat{A} \ket{\Phi_N}^2} - \overline{ \bra{\Phi_N}
\hat{A} \ket{\Phi_N}}^2;
\end{equation}
ii) the \emph{quantum fluctuations} of observable $\hat{A}$ in a single run, quantified by
the observable $(\Delta \hat{A})^2 = (\hat{A}-A_{\Phi_N})^2$. The statistical average of its expectation value in state $\ket{\Phi_N}$ reads
\begin{equation}
\label{2}
\overline{( \Delta A)^2 }  := \overline{ \bra{\Phi_N} \hat{A}^2
\ket{\Phi_N}} - \overline{ \bra{\Phi_N} \hat{A} \ket{\Phi_N}^2}.
\end{equation}
Computations of (\ref{1}) and (\ref{2}) both involve the quartic
average\cite{cumulants} $\overline{z_{\ell_1}^* z_{\ell_2}^*
z_{\ell_3} z_{\ell_4}}$.
Notice, however, that their sum involves only quadratic averages, and is in fact given by the quantum mechanical variance of $\hat{A}$ in the microcanonical state $\varrho_n$ in~(\ref{eq:microcanonical}):
\begin{equation}
\label{totfluc}
\overline{( \Delta A)^2 } + (\delta A)^2  =  ( \Delta A)^2_{\varrho_n}  =
\tr(\hat{\varrho}_n \hat{A}^2 ) - \tr(\hat{\varrho}_n \hat{A} )^2.
\end{equation}

A few comments are in order. If the initial state is sampled from the degenerate distribution with
$n=1$ (which is in fact deterministic and concentrated on the single balanced Fock state), the average
quantum variance $\overline{ (\Delta A)^2 }$ coincides with the
quantum variance of $\ket{\ell=0}$, while, obviously, $(\delta
A)^2=0$. In such a case, the same state is prepared in every run (which requires a very careful preparation procedure and is a somewhat unrealistic assumption for the experiments performed so far).
On the other hand, if the ensemble is made up of eigenstates of
the observable $\hat{A}$, then the quantum fluctuations vanish,
$\overline{ (\Delta A)^2 }=0$, and the only contribution to (\ref{totfluc}) comes
from the statistical fluctuations $(\delta A)^2$, that differentiate individual runs.
An observable is  {\it typical} if $(\delta A)^2 = o
(\overline{A}^2)$, and is {\it stable} at each run if  $\overline{
(\Delta A)^2 } = o (\overline{A}^2)$.

In general, different fluctuations are present in a given
experiment. We analyzed the interference of a two-mode
Bose-Einstein system according to these ideas and showed that some
features of the interference pattern (such as its period and its fringe contrast) are
\textit{robust} against both the afore-mentioned
fluctuations,\cite{typbec} and thus $( \Delta A)^2_{\varrho_n} = o
(\overline{A}^2)$. We shall summarize the main results in the
following sections and shall comment on the role of $( \Delta
A)^2_{\varrho_n}$
for an interesting observable, the 
power spectrum of particle density.\cite{YI}

\section{Observables Related to Interference}
\label{sec-observables}

Since we are interested in the quantities that are related to
interference, we will focus on those observables associated with
the spatial distribution of particles.\cite{ref:PolkovnikovAltmanDemlerPNAS,ref:GritsevAltmanDemlerPolkovnikov-NaturePhys2,ref:PolkovnikovEPL,ref:ImambekovGritsevDemlerVarenna,YI,AYI}
In this section we will review the relevant averages in the
general case, postponing quantitative considerations to the
following section. In the second-quantization formalism, the
spatial density is represented by the operator
\begin{equation}
\hat{\rho}(\bm{r}) = \hat{\Psi}^{\dagger}(\bm{r})
\hat{\Psi}(\bm{r}),
\end{equation}
whose Fourier transform reads
\begin{equation}
\widehat{\tilde{\rho}}(\bm{k}) := \mathcal{F} [\hat{\rho}](\bm{k})
= \int d\bm{r} \, e^{-i \bm{k} \cdot \bm{r} } \hat{\rho}(\bm{r}).
\end{equation}
Expanding the field operators and taking the  expectation value
(\ref{expGEN}), one finds that the average density
\begin{equation}\label{exprho}
 \overline{\rho(\bm{r})} = \overline{ \bra{\Phi_N}
\hat{\rho}(\bm{r}) \ket{\Phi_N}}  = \frac{N}{2}\,\Bigl(
\rho_a(\bm{r}) + \rho_b(\bm{r}) \Bigr), \quad \text{with} \quad
\rho_{a,b}(\bm{r}) := |\psi_{a,b}(\bm{r})|^2,
\end{equation}
is merely the sum of the particle densities in the two modes, with
no interference between them. Clearly, this property holds also
for the Fourier transform. This result apparently contrasts with
experimental observation, as interference is present even if no
phase coherence between the particles in the two modes exists.
However, the average (\ref{exprho}) cannot give sufficient
information on the result of a {\it single} experimental run,
since its fluctuations are very large.

On the other hand, the outcome of a single run can be inferred,
within a controlled degree of approximation, from the study of
the \textit{density power spectrum}, i.e.\ the square modulus of
its Fourier transform:\cite{ref:PolkovnikovEPL,YI,ref:PolkovnikovAltmanDemlerPNAS,ref:GritsevAltmanDemlerPolkovnikov-NaturePhys2,ref:ImambekovGritsevDemlerVarenna,AYI}
\begin{eqnarray}
\label{rho2k}
\hat{R}(\bm{k}) & := &
\widehat{\tilde{\rho}}^\dag(\bm{k})
\widehat{\tilde{\rho}}(\bm{k}) = \hat{r}(\bm{k}) + \hat{N},
\\
\hat{r}(\bm{k}) & := & \int d\bm{r}\, d\bm{r}'
e^{-i\bm{k}\cdot(\bm{r}-\bm{r}')} \hat{\Psi}^{\dagger}(\bm{r})
\hat{\Psi}^{\dagger}(\bm{r}') \hat{\Psi}(\bm{r}')
\hat{\Psi}(\bm{r}) .
\end{eqnarray}
Observe that all states $\ket{\Phi_N}$ in (\ref{psin}) are eigenstates of the total number operator $\hat{N}$ belonging to the eigenvalue $N$, that is fixed, which makes the role of the last addendum in (\ref{rho2k}) immaterial. Notice also that the power spectrum $\hat{R}(\bm{k})$ is
the Fourier transform of the density autocorrelation
function
\begin{equation}
\hat{C}(\bm{r}) = \int d\bm{r}' \hat{\rho}(\bm{r}')
\hat{\rho}(\bm{r}'+\bm{r}).
\end{equation}
Under specific assumptions on the values of $N$ and $n$ in (\ref{6}), we will show that fluctuations around the average value
\begin{equation}
\overline{R(\bm{k})}=\tr\!\left(\hat{\varrho}_n \hat{R}(\bm{k})\right)
\end{equation}
are negligible.

The observable $\hat{r}(\bm{k})$ can be expanded in the mode operators as
\begin{eqnarray}\label{oprk}
\hat{r}(\bm{k}) & = & |\tilde{\rho}_a(\bm{k})|^2 \hat{N}_a
(\hat{N}_a-1) + |\tilde{\rho}_b(\bm{k})|^2 \hat{N}_b (\hat{N}_b-1)
\nonumber\\
&&{}+\Bigl( \tilde{\rho}_a^*(\bm{k}) \tilde{\rho}_b(\bm{k}) +
\tilde{\rho}_b^*(\bm{k}) \tilde{\rho}_a(\bm{k}) +
|\tilde{\rho}_{ba} (\bm{k})|^2 + |\tilde{\rho}_{ab} (\bm{k})|^2
\Bigr) \,\hat{N}_a\hat{N}_b \nonumber \\ &&{} + \Bigl[ \Bigl(
\tilde{\rho}_b(-\bm{k}) \tilde{\rho}_{ba} (\bm{k}) +
\tilde{\rho}_b(\bm{k}) \tilde{\rho}_{ba} (-\bm{k})
\Bigr)\, \hat{N}_{b}{\hat{b}}^{\dagger}\hat{a} \nonumber \\
&&\qquad{} + \Bigl( \tilde{\rho}_{ba} (-\bm{k})
\tilde{\rho}_a(\bm{k}) + \tilde{\rho}_{ba} (\bm{k})
\tilde{\rho}_a(-\bm{k}) \Bigr)\, {\hat{b}}^{\dagger}
\hat{a}\hat{N}_{a} \nonumber \\ &&\qquad{} + \tilde{\rho}_{ba}
(-\bm{k}) \tilde{\rho}_{ba} (\bm{k}) (\hat{b}^{\dagger})^2
{\hat{a}}^2 + \text{h.c.} \Bigr] + \text{other modes},
\end{eqnarray}
with $\tilde{\rho}_{ba}(\bm{k}):=\mathcal{F}[\psi_b^*\psi_a](\bm{k})$. Due to the uniform sampling~(\ref{eq:ensemble}) in $\mathcal{H}_n$,
only the first three operators in \eqref{oprk}, which have
diagonal matrix elements in the Fock basis, yield nonvanishing
contributions to the ensemble average of $\hat{R}(\bm{k})$. The
final result reads
\begin{equation}
\overline{R(\bm{k})} = \overline{r(\bm{k})} +N=
\frac{N^2}{4}\, \Bigl(
|\tilde{\rho}_a(\bm{k})+\tilde{\rho}_b(\bm{k})|^2 +
|\tilde{\rho}_{ba} (\bm{k})|^2 + |\tilde{\rho}_{ab} (\bm{k})|^2
\Bigr) + O(N,n^2), \label{rho2kfinal}
\end{equation}
in which the Fourier transforms of $\psi_b^*\psi_a$, directly
related to interference, appear. [Remember that we always assume $n=o(N)$.]

We now estimate the fluctuations of $\hat R$ according to the philosophy outlined in the previous section. In order to estimate the fluctuations of $\hat{R}(\bm{k})$ around its average and prove that they are small in the large-$N$ limit, we
shall consider the covariance
\begin{eqnarray}\label{DeltaR2}
(\Delta R)^2_{\varrho_n} (\bm{k},\bm{k}') &=& \tr\!\left(
\hat{\varrho}_n \hat{R}(\bm{k}) \hat{R}(\bm{k}') \right) -
\tr\!\left(\hat{\varrho}_n \hat{R}(\bm{k})\right)
\tr\!\left(\hat{\varrho}_n  \hat{R}(\bm{k}') \right)
\nonumber\\
&=&  \tr\!\left(\hat{\varrho}_n \hat{R}(\bm{k}) \hat{R}(\bm{k}') \right) - \overline{r(\bm{k})} \cdot \overline{r(\bm{k}')}.
\end{eqnarray}
It involves (diagonal) matrix elements of the four-particle
correlation function $\prod_{i=1}^4
\hat{\Psi}^{\dagger}(\bm{r}_i) \prod_{i=1}^4
\hat{\Psi}(\bm{r}_i)$.
In the following section, we will analyze
an experimentally relevant case in which the distribution of
$R(\bm{k})$ displays sharp peaks that provide information on the
interference pattern in each experimental run, with fluctuations
being negligible in proper ranges of $N$ and $n$.

\section{Typical Interference of Expanding Gaussian Modes}
\label{sec-gauss}

In the light of the general results on the density power spectrum
$\hat{R}(\bm{k})$ and its fluctuations, we will review in this
section the properties of a realistic model, describing a physical
situation that is close to experimental implementation. The cold
atoms are initially trapped in two Gaussian clouds by an external
double-well potential. The distance between the peaks of the
distributions is significantly larger than their widths, so that
the initial wave packets do not overlap. The trap is then released
and the clouds expand freely until they overlap and interfere. We
will explicitly consider the time evolution of the system in one
spatial dimension, for simplicity: if scattering between the
particles in the condensates is neglected, the time dependence can
be evaluated by observing that the correlation functions at time
$t$ are obtained by replacing the initial modes $\psi_{a,b}(x)$
with their time-evolved counterparts
\begin{equation}
\psi_{a,b}(x,t) = \exp\!\left( \frac{i\hbar t}{2m}
\frac{\partial^2}{\partial x^2} \right) \psi_{a,b}(x),
\end{equation}
with
\begin{equation}
\psi_a(x) = \frac{1}{\pi^{1/4}\sigma^{1/2}} e^{-
(x+\alpha)^2/2\sigma^2}, \qquad \psi_b(x) =
\frac{1}{\pi^{1/4} \sigma^{1/2}} e^{- (x-\alpha)^2/2\sigma^2},
\end{equation}
where $\sigma$ is the width of the Gaussians and $\alpha$ is the half distance between their maxima, chosen large enough in order to ensure that $\psi_{a,b}$ are approximately orthogonal.

A straightforward calculation yields\cite{typbec}
\begin{eqnarray}
\label{DeltaRgauss}
(\Delta R)^2_{\varrho_n} (k,k';t) & = & N^3 C_{3,0}(k,k';t)  +
N n^2 C_{1,2}(k,k';t)  \nonumber \\ & &{} + n^4 C_{0,4}(k,k';t)  +
n^3 C_{0,3}(k,k';t)  + O(N^2),
\end{eqnarray}
where the coefficients $C_{i,j}$ depend on the structure of the modes, but neither on the number of particles $N$, nor on the dimension $n$ of
the sampled Hilbert subspace.

Equation (\ref{DeltaRgauss}) shows that fluctuations are at most $o(N^4)$ when $n=o(N)$. This implies that if $n=o(N)$ (i.e.\ $n/N \to 0$ for $N\to\infty$) fluctuations around the average $\overline{R}$ in
\eqref{DeltaRgauss} in different experimental runs are negligible,
and the distribution of its values is peaked around its most
probable value. These results prove that interference is typical,
and occurs for the overwhelming majority of wave functions of the
condensate when $n=o(N)$.

We observe that similar conclusions are obtained when one deals with plane waves rather than Gaussian modes,\cite{typbec} the only 
difference being in the explicit expression of the coefficients $C_{i,j}$  in Eq.\ (\ref{DeltaRgauss}). Presumably, the dependence on $N$ and $n$ will not change for a wide range of mode functions.
In this sense, our conclusions are of general validity.

\section{The Self-Interacting Gas}
\label{sec-self}

The results reviewed in the previous sections were obtained by averaging over an ensemble of states. It is not immediate to
relate these results to an experiment, since one should introduce a mechanism to sample the random states according
to the desired distribution. We will show in the following that
the randomization of the {\it phases} in the Fock basis emerges in a natural and straightforward way, once interparticle
scattering is considered.

In a classical double-slit experiment, first-order interference is
observable if the relative phase $\phi$ between the incident waves
at each slit does not vary over time. The corresponding case for a
condensate is that of a two-mode {\it coherent} state [see Eq.\ (\ref{coherent}) in the following], in which
all the $N$ particles are in the same superposition of mode wave
functions. A first-order interference pattern can be observed in a
coherent state, with the same offset $\phi$ at each experimental
run.

It is not obvious that a coherent state is created when two (independent) Bose-Einstein condensates are prepared. Nevertheless, we shall now scrutinize the evolution of a coherent state when the
Bose gas is self-interacting. In such a case, the phases in the
Fock basis expansion become (pseudo)random after an initial
transient, which vanishes in the $N\to\infty$ limit. We will show
that this behavior is closely related to a loss of coherence
between the two modes, which leads to the disappearence of the
first-order pattern after the initial transient. This is also
related to the general theory of the Josephson effect\cite{PS} and
spin squeezing.\cite{squeezing} Our previous results ensure, on
the other hand, that an interference pattern can be observed also
in this case, despite its offset fluctuates over time and
experimental runs.

Let us consider a Bose system with particles distributed in two
{\it spatially separated} orthonormal modes $\psi_a$ and $\psi_b$,
whose supports $S_a$ and $S_b$ do not overlap. Assume that
the energies of a single particle in each mode be equal, so that
we can set them equal to zero for convenience, and that the tunneling
between the two modes is negligible. If we also assume that other
modes are made inaccessible (e.g.\ by a large energetic
separation), the two-body contact-interaction term in the Hamiltonian
reads
\begin{eqnarray}\label{pseudorand1}
\hat{H}_\text{int} & = & \frac{g}{2} \int d\bm{r}\,
{\hat{\Psi}}^{\dagger} (\bm{r}) {\hat{\Psi}}^{\dagger} (\bm{r})
\hat{\Psi} (\bm{r}) \hat{\Psi} (\bm{r}) \nonumber \\ & \simeq &
\frac{g}{2} \left( \int_{S_a} d\bm{r}\, {\rho_a^2(\bm{r})} \right)
\hat{N}_a (\hat{N}_a-1) + \frac{g}{2} \left( \int_{S_b} d\bm{r}\,
{\rho_b^2(\bm{r})} \right)\hat{N}_b (\hat{N}_b-1),
\end{eqnarray}
where the coupling constant $g$ is determined at the lowest order
by the scattering length $a_s$ through $g=4\pi\hbar^2 a_s/m$, and
the products of $a$ and $b$ mode operators do not
appear because the supports of the modes have no
overlap. If the integrals appearing in \eqref{pseudorand1} are
equal (e.g.\ if the mode density profiles are related by
translation and/or reflection), the Hamiltonian reduces to
\begin{eqnarray}
\hat{H}_\text{int} & = & \frac{\tilde{g}}{2} \left[ {\hat{N}_a}^2 +
{\hat{N}_b}^2 - (\hat{N}_a + \hat{N}_a) \right] \nonumber \\ & = &
\frac{\tilde{g}}{2} \left[ \frac{(\hat{N}_a - \hat{N}_b)^2}{2} +
\frac{(\hat{N}_a + \hat{N}_b)^2}{2} - (\hat{N}_a + \hat{N}_a)
\right],
\end{eqnarray}
where $\tilde{g} := g \int_{S_{a}} d\bm{r}\,
{\rho_{a}^2(\bm{r})}$. Since $\hat{N}_a+\hat{N}_b= \hat{N}$ is a constant of
motion and we are going to consider states with a fixed number of
particles $N$, distributed among the two modes, the only part of the
Hamiltonian which is relevant to the evolution of a state
$\ket{\Phi_N}$ is
\begin{equation}\label{hdelta}
\hat{h} = \tilde{g} ( \delta \hat{N} )^2,\quad
\text{with} \quad \delta \hat{N} = \frac{\hat{N}_a-\hat{N}_b}{2}.
\end{equation}
By definition,  the $N$-particle two-mode Fock basis~(\ref{fock}) satisfies
\begin{equation}
\delta \hat{N} \ket{\ell} = \ell \ket{\ell}.
\end{equation}

\section{Dynamics and Typicality}
\label{sec-typs}
Let us consider an initial coherent state,
\begin{equation}
\label{coherent}
\ket{\Psi_0} = \frac{1}{\sqrt{N!}} \left( \frac{\hat{a}^{\dagger}
+ \hat{b}^{\dagger}}{\sqrt{2}} \right)^N
\ket{\Omega} = \frac{1}{2^{N/2}} \sum_{\ell=-N/2}^{N/2} \left( \begin{array}{c} N \\
N/2+\ell
\end{array}\right)^{\frac{1}{2}} \ket{\ell},
\end{equation}
in which all the particles are created in the wave function
$[\psi_a(\bm{r}) + \psi_b(\bm{r})]/\sqrt{2}$. Once
the initial wave packets are let to expand and \textit{overlap}, this state,
along with all the states
\begin{equation}
\label{coherentphi}
\ket{\varphi} = \frac{1}{\sqrt{N!}} \left( \frac{e^{i\varphi/2}
\hat{a}^{\dagger} + e^{-i\varphi/2} \hat{b}^{\dagger}}{\sqrt{2}}
\right)^N \ket{\Omega} = \frac{1}{2^{N/2}} \sum_{\ell=-N/2}^{N/2}
\left(
\begin{array}{c} N \\ N/2+\ell \end{array}\right)^{\frac{1}{2}}
e^{i \ell \varphi} \ket{\ell},
\end{equation}
displays first-order interference, appearing in the expectation
value $\bra{\varphi}{\hat{\rho}}(\bm{r},t)\ket{\varphi}$ with
 maximal visibility, since the relative phase between the two modes is fixed and the modes are equally populated on
average.
The coherent states (\ref{coherentphi}), called phase states,\cite{Leggettbook,Molmer,CD,ref:LeggettSols} which are relevant to describe
interference, do not form a basis. Their overlap reads
\begin{equation}
\bra{\varphi} \varphi'\rangle = \left( \cos
\frac{\varphi-\varphi'}{2} \right)^N
\end{equation}
and is characterized for large $N$ by a sharp peak around
$\varphi-\varphi'=0$, whose width is~$O(N^{-1/2})$.

Even if all the coefficients of the states \eqref{coherent}--\eqref{coherentphi} in the Fock basis are
nonvanishing, the presence of the binomial coefficient implies
that for large $N$ the states can be very well approximated by
truncating the sum at $|\ell|<\ell_{\max}=O(N^{1/2})$:
the approximate states thus belong to $\mathcal{H}_{2\ell_\text{max}}$ [see Eq.\ (\ref{6})].
The evolution of the initial state (\ref{coherent}) generated by the
Hamiltonian \eqref{hdelta} reads
\begin{equation}\label{coherentevo}
\ket{\Psi(t)}
=\sum_{\ell=-N/2}^{N/2} \left(
\begin{array}{c} N \\ N/2+\ell \end{array}\right)^{\frac{1}{2}}
\frac{e^{-i \ell^2 \tilde{g} t}}{2^{N/2}} \ket{\ell}.
\end{equation}
In general, the evolved state at $t>0$ is no longer a coherent
state of the form \eqref{coherentphi}: the delicate phase relation
between the amplitudes in the Fock basis breaks due to the
time-dependent phase factors, which are quadratic in the imbalance~$\ell$. This behavior produces a pseudo-randomization for
irrational values of $\tilde{g}t/\pi$, which simulates the random
phase sampling in the statistical ensemble for a fixed
distribution of the imbalances. A detailed analysis of this aspect will be presented in the following.

\begin{figure}[t]
\centering
\includegraphics[width=0.6\textwidth]{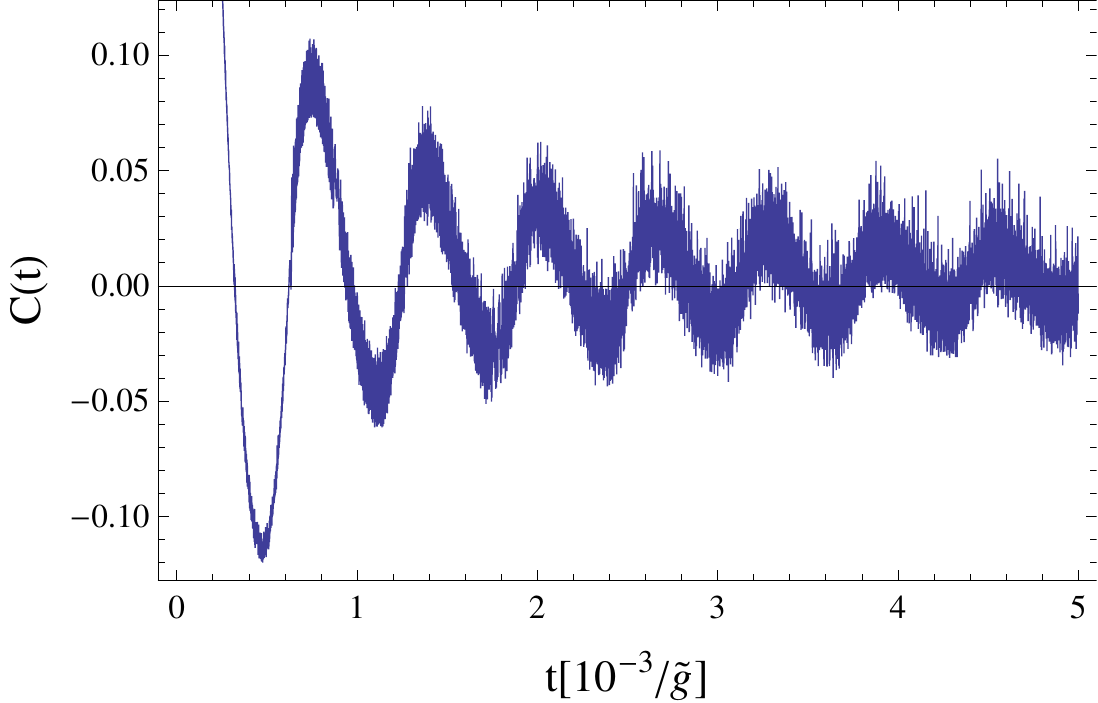}
\caption{Autocorrelation $C(t)$, defined in (\ref{fl}) and
(\ref{auto}), of the sequence of phases $f_{\ell}(t)$ in the case
$N=10^4$. Though very rapidly variating, the function is analytic
for all finite $N$. The spacing between points in the horizontal
axis is $t=10^{-6}/\tilde{g}$. Recall that at the initial time the
sequence of coefficients is perfectly correlated [$C(0)=1$, not
shown in the plot]. }\label{autocorr}
\end{figure}

The loss of relative coherence can be quantified by
studying the squared component of the state $\ket{\Psi(t)}$ along
the general coherent state $\ket{\varphi}$:
\begin{equation}\label{projection}
P_{\varphi}(t) = |\bra{\varphi} \Psi(t)\rangle|^2 = \left|
\frac{1}{2^N} \sum_{\ell=-N/2}^{N/2} \left(
\begin{array}{c} N \\ N/2+\ell \end{array}\right)e^{-i
\ell \varphi - i \ell^2 \tilde{g} t}\right|^2.
\end{equation}
A peaked distribution around one value of the relative phase
$\varphi$ indicates a high degree of coherence, and thus a large
visibility in the first-order interference. On the other hand, an
almost uniform value of \eqref{projection} in $(-\pi,\pi]$
indicates that very small phase coherence between the
particles in the two modes is present, and the offset of the
interference fringes randomly fluctuates from run
to run, leading to the disappearance of first-order
interference effects. The relative phase distribution of the
initial state \eqref{coherent} is peaked around $\varphi=0$.
Notice that the peak has a finite width, since each
coherent state can be expressed as a linear superposition of the
others. The peak becomes sharper as $N$ increases, with a standard
deviation
\begin{equation}
\sigma_\varphi(0)=\sqrt{\int_{-\pi}^{+\pi}d\varphi\, \varphi^2P_\varphi(0)\Big/\int_{-\pi}^{+\pi}d\varphi \,P_\varphi(0)}=\sqrt{2\over N}.
\end{equation}
As the system evolves, the initial value of $\sigma_\varphi$ becomes
negligible, as the standard deviation increases linearly in time,
like
\begin{equation}
\sigma_{\varphi} (t) \simeq \sqrt{N/2}\, \tilde{g} t.
\end{equation}
This result can be immediately obtained by a Gaussian
approximation of the binomial coefficients,
\begin{equation}
{1\over2^N}\left(\begin{array}{c} N \\ N/2+\ell \end{array}\right)\sim\sqrt{2\over\pi N}e^{-{2\over N}\ell^2},
\end{equation}
for $N\to \infty$. At  time $t_* =2 \pi/\tilde g\sqrt{2 N}$, one
expects that the state is spread over all possible values of the
relative phase. This does not prevent coherence to be recovered at a subsequent time: indeed the state is again perfectly coherent at
$t=\pi/\tilde{g}$, because $\ket{\Psi(\pi/\tilde{g})}=\ket{\varphi=\pi}$, and returns to the
initial state $\ket{\varphi=0}$ after the recurrence time
$t=t_r:=2\pi/\tilde{g}$.

\section{Phase Randomization}
\label{sec-pseudo}

\begin{figure}
\begin{center}
\begin{tabular}{cc}
\multicolumn{2}{c}{\footnotesize(a) $\tilde{g}t=10^{-6}$}\\[1mm]
\includegraphics[width=0.42\textwidth]{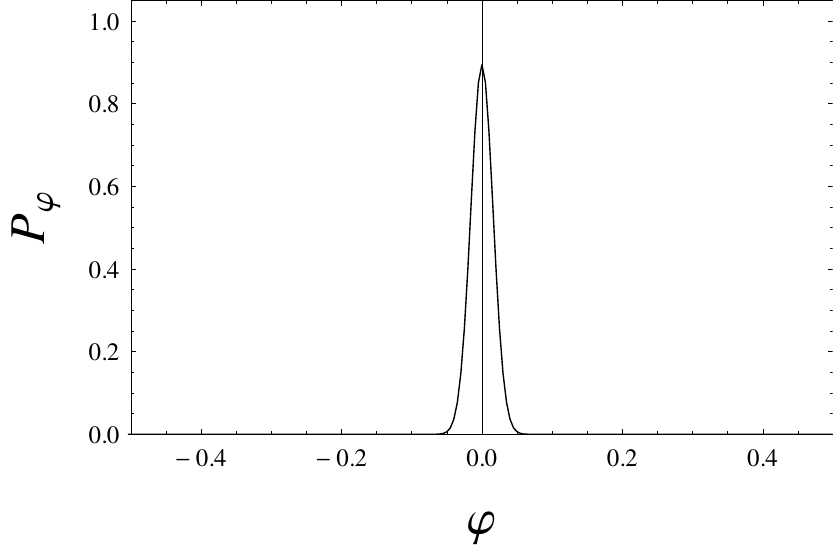}&
\includegraphics[width=0.42\textwidth]{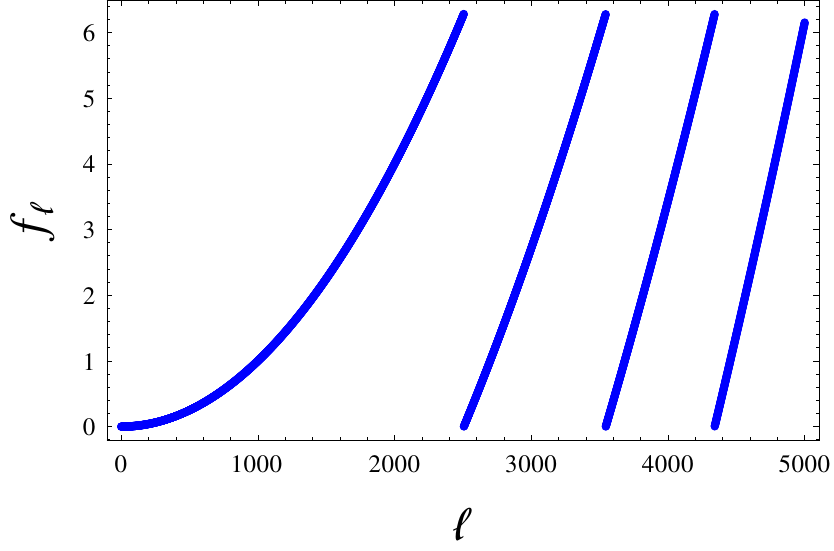}\\
\multicolumn{2}{c}{\footnotesize(b) $\tilde{g}t={0.044}$}\\
\includegraphics[width=0.42\textwidth]{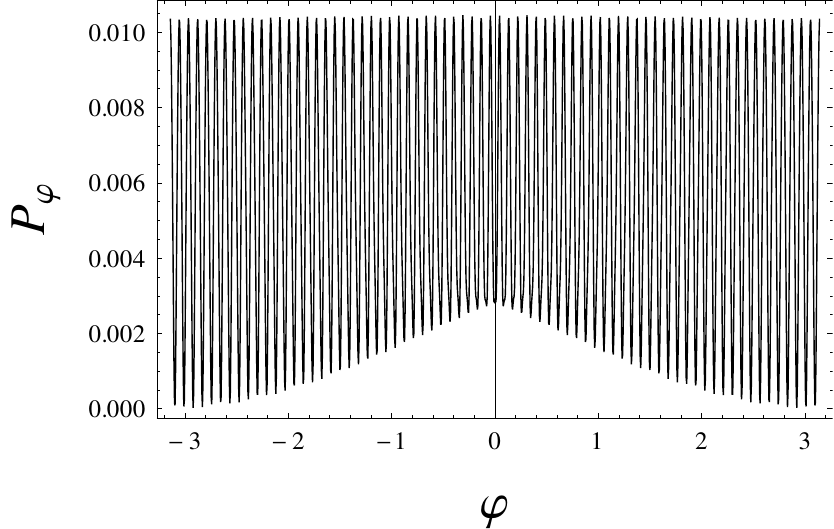}&
\includegraphics[width=0.42\textwidth]{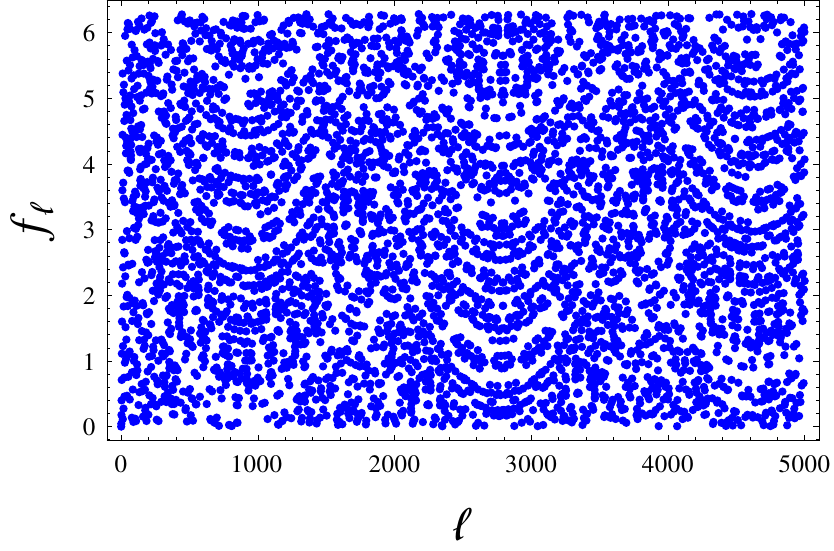}\\
\multicolumn{2}{c}{\footnotesize(c) $\tilde{g}t=1.5$}\\
\includegraphics[width=0.42\textwidth]{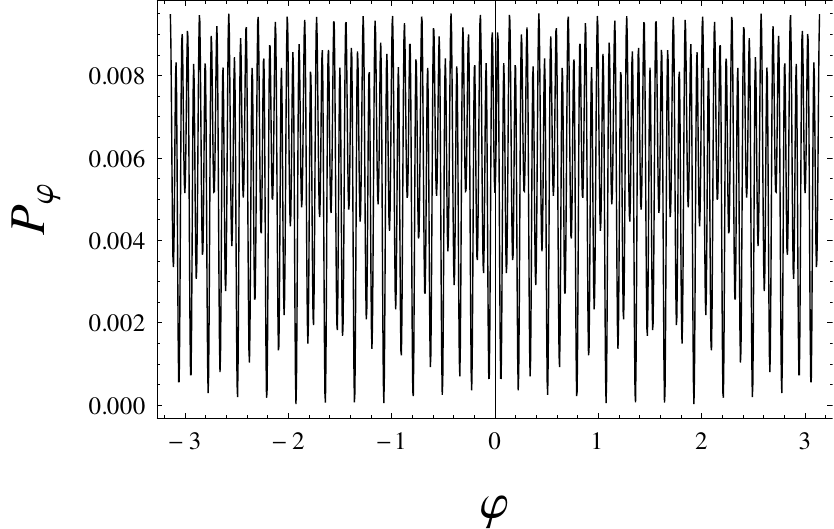}&
\includegraphics[width=0.42\textwidth]{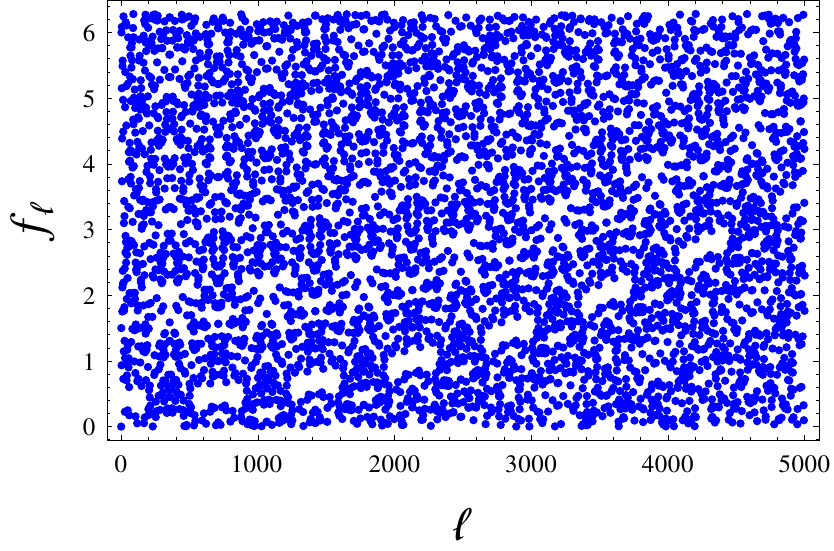}\\
\multicolumn{2}{c}{\footnotesize(d) $\tilde{g}t=5\pi/7$}\\
\includegraphics[width=0.42\textwidth]{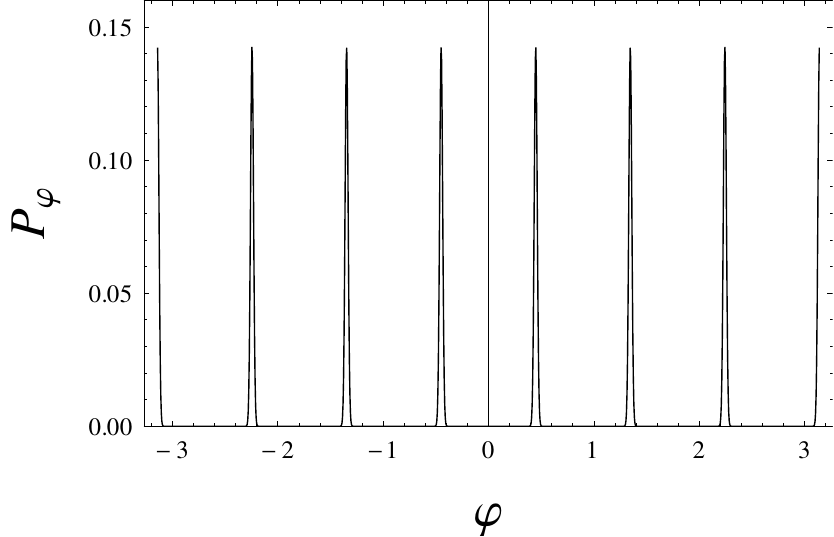}&
\includegraphics[width=0.42\textwidth]{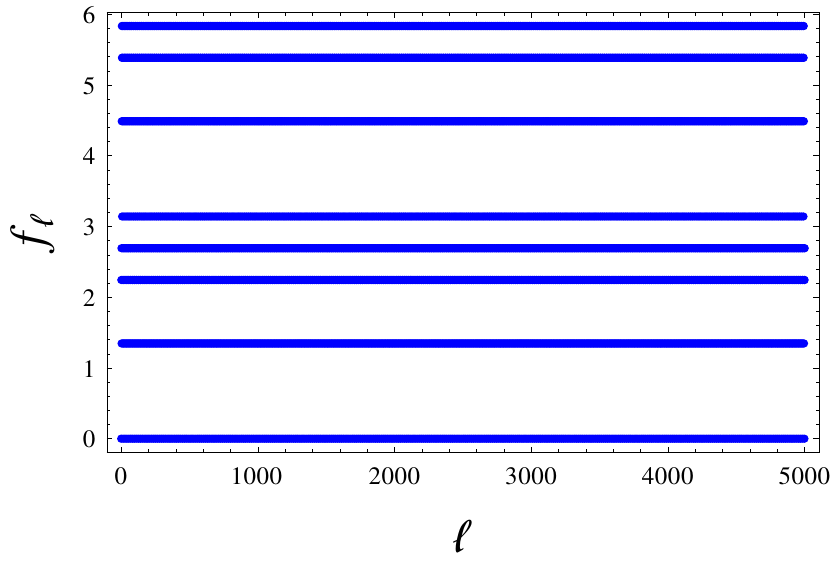}
\end{tabular}
\end{center}
\caption{ The two columns display in parallel, for different times,
the behavior of the squared component $P_{\varphi}(t)$ of
$\ket{\Psi(t)}$ along the coherent state defined by phase $\varphi$ (left panels), and the distribution of the phases $f_{\ell}(t)$ of the
coefficients in the Fock basis (right panels) [see \eqref{projection}--\eqref{fl}]. The loss of the initial coherence
from (a) to (b) is due to the randomization of the phases. The
last two plots refer to long-time cases in which (c) there is no
coherence and (d) partial coherence is
recovered.}\label{fig:evolution}
\end{figure}

Let us now discuss the phase randomization process
which involves the coefficients of $\ket{\Psi(t)}$ in the
expansion \eqref{coherentevo}: their phases are 
\begin{equation}\label{fl}
f_{\ell}(t) := \tilde{g} t \ell^2 \mod 2\pi.
\end{equation}
Can the sequence $\{f_{\ell}\}_{|\ell|\leq N/2}$ mimic a random
sequence, sampled from a uniform distribution in $[0,2\pi]$, for
some time $t$? To address this question, one can analyze a
quantity which is common in testing pseudorandom numbers, namely
the autocorrelation between  nearest phases
\begin{equation}
\label{auto} C(t) = \frac{ (N+1) \sum_{\ell} f_{\ell}(t)
f_{\ell+1}(t) - [\sum_{\ell} f_{\ell}(t)]^2 }{ (N+1) \sum_{\ell}
f_{\ell}^2(t)- [\sum_{\ell} f_{\ell}(t)]^2},
\end{equation}
which is expected to vanish for a truly random sequence of independent phases. At the
initial time, the values of the phases for adjacent $\ell$'s are
strongly correlated, since $C(0)=1$.
The autocorrelation reduces
as time increases: for $N=10^4$, its value typically drops down to
$|C(t)|\lesssim 0.05$ at $\tilde{g}t\simeq 2\cdot 10^{-3}$, with
oscillations around zero (see Fig.\ \ref{autocorr}). An exception
to this behavior occurs when $\tilde{g}t/\pi$ is a rational
number: in this case, phases are not uniformly distributed in
$[0,2\pi]$, since they can assume only a finite set of values.
In Fig.\ \ref{fig:evolution}, the projection $P_{\varphi}(t)$ of
the state $\ket{\Psi(t)}$ on the coherent state with relative phase $\varphi$ [see
\eqref{projection}] is plotted as a function of $\varphi$, in
parallel with a scatter plot of the phases $f_{\ell}(t)$. 

The plot in Fig.\ \ref{fig:evolution}(a) shows
a distribution $P_{\varphi}$ evaluated at $\tilde{g}t=10^{-6}$,
which is still peaked around $\varphi=0$, indicating a good degree
of coherence (left panel). On the other hand, the phases $f_{\ell}$ of adjacent
Fock coefficients are manifestly correlated, with
$C(10^{-6}/\tilde{g})=0.997$ (right panel).

At time $\tilde{g}t_* =\sqrt{2/N}\,\pi \simeq 0.044$,
the function $P_{\varphi}$ spreads over the whole interval
$(-\pi,\pi]$, as shown in Fig.\ \ref{fig:evolution}(b) (left panel). No correlations manifestly 
emerge between the phases of the coefficients, which appear to be (almost)
uniformly scattered (right panel). This observation is confirmed by the small
value of the autocorrelation $C(0.044/\tilde{g})= -3.4\cdot 10^{-3}$ (not shown in Fig.\ \ref{autocorr}, where shorter times $t \leq 5 \cdot 10^{-3}/\tilde g$ are displayed). 

In Figs.\ \ref{fig:evolution}(c) and \ref{fig:evolution}(d), two different situations that can occur at large times are presented (both cases are not shown in Fig.\ \ref{autocorr}, that displays times of order $t \sim 10^{-3}/\tilde g$).
When $\tilde{g}t/\pi$ is irrational, Fig.\ \ref{fig:evolution}(c), the phases of the
coefficients are pseudorandom [$C(1.5/\tilde{g})\simeq 10^{-2}$] (left panel),
and the relative phase distribution tends to fill the interval
$(-\pi,\pi]$ uniformly (right panel). On the other hand, when
$\tilde{g}t/\pi=p/q$ is rational, Fig.\ \ref{fig:evolution}(d), the phases of the coefficients
become periodic, with period $q$ if $p$ is even and $2q$ if $p$ is
odd: no randomization occurs in this case, and the state of the
system appears as a superposition of a finite number of coherent
states (left panel). The phases are very correlated (right panel).

Let us finally remark that full correlation among the coefficients
is recovered, together with coherence, when $t=t_r/2$. Our case
study shows how the randomization of the coefficients and the loss of
coherence between the two modes are strictly correlated. If one
observes the state of the system at a generic time $t$ in
$[0,t_r]$, in the vast majority of cases the phases of the
coefficients behave as if they were sampled from a random
distribution. One can thus expect in this case that the typical
properties of the system are captured by the average over a
distribution of states with the random phases of the Fock basis
coefficients.

It should be observed that in the case here considered there is
no mechanism that yields a randomization of the {\it amplitudes}
of the coefficients. Of course, it is possible to conceive several
physical situations that do not preserve the imbalance
distribution $\ell$, and  randomize also the amplitudes. However,
the results of the previous section crucially depend only on the
randomness of the phases, and seem to indicate that one should
expect only slightly quantitative differences if the amplitudes
were sampled differently from the uniform
sampling~(\ref{eq:ensemble}).

\section{Conclusions}
\label{sec-concl}

Typicality is a fecund concept in modern statistical mechanics.
Typical phenomena characterize physical situations with overwhelming probability. We have shown in Ref.\ \refcite{typbec} and in this article that the interference of two \emph{independently prepared}.  BECs is typical, namely (almost) always occurs
when two BECs are created and let to interfere. This interference is not of first order.  
We therefore looked at the relative phase of the condensates in order to elucidate its randomization
mechanism.

We showed that self-interaction (accounting for two-body scattering processes) within the condensates yields such phase randomization and makes first-order interference vanish. After a certain time, that is inversely proportional to the interaction strength, the relative phase is randomized. However, an interference pattern
will still be observed, as a consequence of typicality. The interplay between the absence of first-order interference and the (observable and experimentally observed) presence of second-order interference is an interesting phenomenon, that characterizes the physics of BECs.


\begin{thebibliography}{10}

\bibitem{PW}
A.~S. Parkins and D.~F. Walls, \textit{Phys. Rep.} \textbf{303}  (1998) 1.

\bibitem{SB}
F. Dalfovo, S. Giorgini, L.~P. Pitaevskii and S. Stringari, \textit{Rev. Mod.
  Phys.} \textbf{71}  (1999) 463.

\bibitem{Leggett}
A.~J. Leggett, \textit{Rev. Mod. Phys.} \textbf{73}  (2001) 307.

\bibitem{BDZ}
I. Bloch, J. Dalibard and W. Zwerger, \textit{Rev. Mod. Phys.} \textbf{80}
  (2008) 885.

\bibitem{PS}
L. Pitaevskii and S. Stringari, \textit{Bose-Einstein Condensation} (Oxford
  University Press, Oxford, 2003).

\bibitem{PeSm}
C.~J. Pethick and H. Smith, \textit{Bose-Einstein Condensation in Dilute
  Gases}, 2nd edn. (Cambridge University Press, Cambridge, 2008).

\bibitem{Leggettbook}
A.~J. Leggett, \textit{Quantum Liquids: Bose Condensation and Cooper Pairing in
  Condensed-Matter Systems} (Oxford University Press, Oxford, 2006).

\bibitem{exptBEC}
M.~R. Andrews, C.~G. Townsend, H.-J. Miesner, D.~S. Durfee, D.~M. Kurn and W.
  Ketterle, \textit{Science} \textbf{275}  (1997) 637.

\bibitem{Dirac}
P.~A.~M. Dirac, \textit{The Principles of Quantum Mechanics}, 4th edn. (Oxford
  University Press, Oxford, 1958).

\bibitem{feynman}
R. Feynman, R. Leighton and M. Sand, \textit{The Feynman Lectures on Physics}
  (Addison Wesley Longman, Reading, MA, 1970), Vol.~3.

\bibitem{NP93}
M. Namiki and S. Pascazio, \textit{Phys. Rep.} \textbf{232}  (1993) 301.

\bibitem{NPN}
M. Namiki, S. Pascazio and H. Nakazato, \textit{Decoherence and Quantum
  Measurements} (World Scientific, Singapore, 1997).

\bibitem{HBT1}
R. {Hanbury Brown} and R.~Q. Twiss, \textit{Nature (London)} \textbf{177}
  (1956) 27.

\bibitem{HBT}
R. {Hanbury Brown} and R.~Q. Twiss, \textit{Nature (London)} \textbf{178}
  (1956) 1046.

\bibitem{ref:MandelWolf}
L. Mandel and E. Wolf, \textit{Optical Coherence and Quantum Optics} (Cambridge
  University Press, Cambridge, 1995).

\bibitem{JY}
J. Javanainen and S.~M. Yoo, \textit{Phys. Rev. Lett.} \textbf{76}  (1996) 161.

\bibitem{CGNZ}
J.~I. Cirac, C.~W. Gardiner, M. Naraschewski and P. Zoller, \textit{Phys. Rev.
  A} \textbf{54}  (1996) 3714(R).

\bibitem{WCW}
T. Wong, M.~J. Collett and D.~F. Walls, \textit{Phys. Rev. A} \textbf{54}
  (1996) 3718(R).

\bibitem{Molmer}
K. M\o lmer, \textit{Phys. Rev. A} \textbf{55} (1997) 3195.

\bibitem{CD}
Y. Castin and J. Dalibard, \textit{Phys. Rev. A} \textbf{55}  (1997) 4330.

\bibitem{Raz}
K. G\'oral, M. Gajda and K. Rza\ifmmode \mbox{\c{}}\else
  \c{}\fi{}\ifmmode~\dot{z}\else \.{z}\fi{}ewski, \textit{Phys. Rev. A}
  \textbf{66}  (2002) 051602(R).

\bibitem{Benatti1}
S. Anderloni, F. Benatti, R. Floreanini and A. Trombettoni, {\it
J. Phys. A: Math. Theor.} {\bf 42} (2009) 035306.

\bibitem{Benatti2}
F. Benatti, R. Floreanini,  and G.~G. Guerreschi, {\it Phys. Lett. A}
{\bf 373} (2009) 3516. 

\bibitem{ref:LeggettSols}
A.~J. Leggett and F. Sols, \textit{Found. Phys.} \textbf{21}  (1991) 353.

\bibitem{typbec}
P. Facchi, H. Nakazato, S. Pascazio, F.~V. Pepe and K. Yuasa, \textit{Phys.
  Rev. A} \textbf{89}  (2014) 063625.

\bibitem{Schmied}
T. Schumm, S. Hofferberth, L.~M. Andersson, S. Wildermuth, S. Groth, I.
  Bar-Joseph, J. Schmiedmayer and P. Kr\"uger, \textit{Nature Phys.} \textbf{1}
   (2005) 57.

\bibitem{Schmied2}
S. Hofferberth, I. Lesanovsky, B. Fischer, J. Verdu and J. Schmiedmayer,
  \textit{Nature Phys.} \textbf{2}  (2006) 710.

\bibitem{Schmied3}
S. Hofferberth, I. Lesanovsky, B. Fischer, T. Schumm and J. Schmiedmayer,
  \textit{Nature (London)} \textbf{449}  (2007) 324.

\bibitem{Schmied4}
S. Hofferberth, I. Lesanovsky, T. Schumm, A. Imambekov, V. Gritsev, E. Demler
  and J. Schmiedmayer, \textit{Nature Phys.} \textbf{4}  (2008) 489.

\bibitem{Schmied5}
T. Langen, R. Geiger, M. Kuhnert, B. Rauer and J. Schmiedmayer, \textit{Nature
  Phys.} \textbf{9}  (2013) 640.

\bibitem{ref:PolkovnikovEPL}
A. Polkovnikov, \textit{Europhys. Lett.} \textbf{78}  (2007) 10006.

\bibitem{ref:Paraoanu-JLTP}
G.~S. Paraoanu, \textit{J. Low Temp. Phys.} \textbf{153}  (2008) 285.

\bibitem{ref:Paraoanu-PRA}
G.~S. Paraoanu, \textit{Phys. Rev. A} \textbf{77}  (2008) 041605(R).

\bibitem{YI}
M. Iazzi and K. Yuasa, \textit{Phys. Rev. A} \textbf{83}  (2011) 033611.

\bibitem{esteve}
K. Maussang, G.~E. Marti, T. Schneider, P. Treutlein, Y. Li, A. Sinatra, R.
  Long, J. Est\`eve and J. Reichel, \textit{Phys. Rev. Lett.} \textbf{105}
  (2010) 080403.

\bibitem{cumulants}
P. Facchi, G. Florio, U. Marzolino, G. Parisi and S. Pascazio, \textit{J. Phys.
  A: Math. Theor.} \textbf{43}  (2010) 225303.

\bibitem{ref:PolkovnikovAltmanDemlerPNAS}
A. Polkovnikov, E. Altman and E. Demler, \textit{Proc. Natl. Acad. Sci. USA}
  \textbf{103}  (2006) 6125.

\bibitem{ref:GritsevAltmanDemlerPolkovnikov-NaturePhys2}
V. Gritsev, E. Altman, E. Demler and A. Polkovnikov, \textit{Nature Phys.}
  \textbf{2}  (2006) 705.

\bibitem{ref:ImambekovGritsevDemlerVarenna}
A. Imambekov, V. Gritsev and E. Demler,  in \textit{Ultra-Cold Fermi Gases},
  Vol.~164 of \textit{International School of Physics ``Enrico Fermi''}, ed.~M.
  Inguscio, W. Ketterle and C. Salomon (IOS, Amsterdam, 2007), pp.\ 535--606.

\bibitem{AYI}
S. Ando, K. Yuasa and M. Iazzi, \textit{Int. J. Quant. Inf.} \textbf{9}  (2011)
  431.

\bibitem{squeezing}
J. Ma, X. Wang, C. Sun and F. Nori, \textit{Phys. Rep.} \textbf{509}
  (2011) 89.

\end{thebibliography}
\end{document}